\documentclass[10pt,conference]{IEEEtran}
\IEEEoverridecommandlockouts

\usepackage{cite}
\usepackage{amsmath,amssymb,amsfonts}
\usepackage{algorithmic}
\usepackage{graphicx}
\usepackage{textcomp}
\usepackage{colortbl}  
\usepackage{booktabs}
\usepackage{multirow}
\usepackage{xcolor}

\usepackage[switch]{lineno}
\usepackage{hyperref}
\usepackage{algorithm}
\usepackage{algorithmic} 
\usepackage{tcolorbox}

\usepackage{titlesec} 
\titlespacing{\section}{0pt}{*1}{*1}

\def\BibTeX{{\rm B\kern-.05em{\sc i\kern-.025em b}\kern-.08em
    T\kern-.1667em\lower.7ex\hbox{E}\kern-.125emX}}
\begin{document}

\title{XBRLTagRec: Domain-Specific Fine-Tuning and Zero-Shot Re-Ranking with LLMs for Extreme Financial Numeral Labeling}

\author{
     \IEEEauthorblockN{Gang Hu, Qun Zhang, Jingyao Luo, Yile Jiang, Jing Chai, Haiyan Ding\thanks{*Corresponding Author: dinghaiyan@ynu.edu.cn}}
     \vspace{5pt}
     \IEEEauthorblockA{School of Information Science \& Engineering, Yunnan University, Kunming, China}
}


\maketitle

\begin{abstract}
Publicly traded companies must disclose financial information under regulations of the Securities and Exchange Commission (SEC) and the Generally Accepted Accounting Principles (GAAP). The eXtensible Business Reporting Language (XBRL), as an XML-based financial language, enables standardized and machine-readable reporting, but accurate tag selection from large taxonomies remains challenging. Existing fine-tuning-based methods struggle to distinguish highly similar XBRL tags, limiting performance in financial data matching.
To address these issues, we introduce XBRLTagRec\footnote{\url{https://github.com/ckw29/XBRLTagRec}}, an end-to-end framework for automated financial numeral tagging. The framework generates semantic tag documents with a fine-tuned FLAN-T5-Large model, retrieves relevant candidates via semantic similarity, and applies zero-shot re-ranking with ChatGPT-3.5 to select the optimal tag. Experiments on the FNXL dataset show that XBRLTagRec outperforms the state-of-the-art FLAN-FinXC framework, achieving 2.64\%–4.47\% improvements in Hits@1 and Macro metrics. These results demonstrate its effectiveness in large-scale and semantically complex tag matching scenarios.

\end{abstract}


\section{Introduction}


Publicly listed companies are required to disclose standardized financial information in compliance with regulatory frameworks like GAAP~\cite{enriques2015disclosure}. XBRL, an XML-based standard, is widely adopted to enable structured, machine-readable reporting, improving the interpretability and comparability of financial data~\cite{richards2006introduction}. Accurately assigning concept tags in complex financial texts remains a key challenge.


In practice, XBRL tag selection remains largely manual, requiring substantial domain expertise and contextual understanding~\cite{troshani2010translation}. The large scale of XBRL taxonomies, semantic overlap among tag definitions, and inconsistent financial language usage make automated tag matching highly challenging. These issues lead to frequent annotation errors and limit the scalability of intelligent financial data processing.

Prior work has explored automated financial numeral labeling from multiple perspectives. FiNER~\cite{loukas2022finer} formulated XBRL tag assignment as a NER task using BERT-based sequence labeling, but its limited tag coverage restricts real-world applicability. Subsequent studies applied extreme multi-label classification on the FNXL dataset~\cite{sharma2023financial}, yet such supervised methods struggle to distinguish highly similar tags and fail to fully exploit the semantic structure of XBRL taxonomies. More recent approaches incorporate tag semantics via graph-based models or generative frameworks~\cite{saini2021galaxc,ma2022label,khatuya2024parameter}, while fine-tuned financial LLMs mainly focus on entity-centric NER rather than financial numerals~\cite{hu2024no,wang2025fintagging}. Consequently, accurate semantic differentiation and ranking of extremely similar XBRL numeral tags remains an open challenge.

\begin{figure}[h]
    \centering
    \includegraphics[width=3.5in]{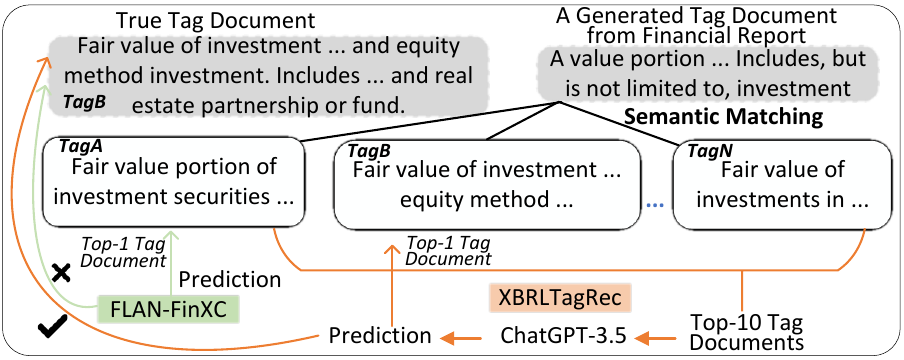}
    \vspace{-21pt} 
    \caption{Illustrates the effectiveness of ChatGPT-3.5 in semantic re-ranking extremely similar label documents.}
    \label{fig:case}
\end{figure}

Despite the progress made by these methods from various perspectives, they still face three main challenges in practical applications: (1) In ultra-large tag systems, NER and classification methods struggle to effectively differentiate semantically similar tags; (2) Existing methods generally rely on static tag encoding, which fails to fully leverage the rich semantic features of XBRL tag documents; (3) While generative methods can provide candidate label documents, they still lack robust and efficient semantic ranking mechanisms to filter the optimal results. These issues collectively lead to performance bottlenecks in current models. As shown in Figure~\ref{fig:case}, FLAN-FinXC is an effective state-of-the-art baseline for financial numeral labeling. However, it may still miss the correct label within its Top-k matches when relying on Top-1 matching. In contrast, the introduction of ChatGPT, with its broad financial knowledge and strong learning capabilities, enables iterative differentiation of these highly challenging, extremely similar labels, leading to more accurate label predictions.

\vspace{-12pt} 
\vspace{-9pt}


\begin{figure*}[h]
    \centering
    \includegraphics[width=7.05in]{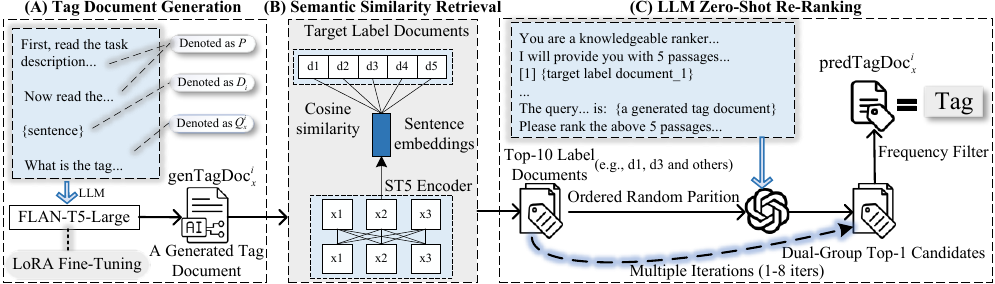}
    \vspace{-10pt} 
    \caption{Our proposed XBRLTagRec framework consists of three components: (A) Tag Document Generation, (B) Semantic Similarity Retrieval, and (C) LLM Zero-Shot Re-Ranking. First, a FLAN-T5 LoRA-tuned LLM generates a tag document from instruction prompts, financial text, and related numeric questions. Next, the Sentence-T5-XXL model encodes this document and all ground-truth XBRL documents to retrieve the Top-10 most similar ones via cosine similarity. Finally, ChatGPT-3.5 performs multi-round re-ranking of these candidates, and the final XBRL tags are selected by majority vote.}
    \label{fig:overview}
\vspace{-1.5em}
\end{figure*}

In this paper, we propose XBRLTagRec, an end-to-end framework for automated financial numeral labeling that matches target numbers in financial texts to standardized XBRL tags. XBRLTagRec leverages instruction-tuned LLMs to generate label documents and integrates semantic retrieval with zero-shot re-ranking to select the optimal tag from a large candidate set, improving both accuracy and interpretability. XBRLTagRec consists of three modules:
(1) Tag Document Generation, where FLAN-T5-Large is fine-tuned with LoRA to generate semantic label documents from financial text, prompts, and target numbers;
(2) Semantic Similarity Retrieval, which uses Sentence-T5-XXL to retrieve the Top-10 most relevant XBRL label documents;
(3) LLM Zero-Shot Re-Ranking, where ChatGPT performs multi-round semantic re-ranking with frequency voting to determine the final label.
This design forms a closed-loop pipeline from generation to selection and effectively exploits the rich semantics of XBRL tag documents in large-scale, highly similar label spaces.


To evaluate effectiveness, we conduct experiments on a subset of the FNXL dataset. XBRLTagRec outperforms the state-of-the-art baseline FLAN-FinXC, achieving improvements of 2.64\%–4.47\% in Hits@1 and Macro metrics, showing strong robustness in accurate prediction and multi-label balance.

In summary, the main contributions are as follows:
\textbf{(1) We propose an XBRL tag prediction framework, XBRLTagRec}, which combines the generative capabilities of LLMs, semantic similarity retrieval, and multi-round iterative re-ranking mechanism. It builds an end-to-end pipeline solution that enhances the accuracy of matching financial text numbers with XBRL tags.
\textbf{(2) We introduce an iterative re-ranking method via filtering and frequency voting}, leveraging the semantic reasoning capabilities and statistical properties of ChatGPT-3.5. This approach achieves robust ranking of candidate label documents, significantly improving the stability and semantic alignment of predicted labels.
\textbf{(3) We develop a modular and highly pluggable framework architecture}, with clear boundaries between each module. This modular structure allows for flexible replacement of generators, encoders, and re-rankers, thus supporting the seamless integration of different LLMs and inference strategies. This flexibility enhances the system's generalizability and engineering adaptability.
 
\section{RELATED WORK}
\textbf{(1) XBRL Tagging Methods:} To improve the accuracy of XBRL tag annotation for numerical data in financial texts, early studies used structured knowledge and domain-specific modeling techniques. \cite{saini2021galaxc} proposed GalaXC, which built a graph neural network based on hierarchical relationships among tags to enhance the model’s understanding of semantic structures. \cite{sharma2023financial} released the FNXL dataset and used a pipeline-based approach. \cite{wang2023standardizing} integrated various semantic similarity metrics to align custom tags with standard taxonomies.
With the rapid progress of LLMs, researchers began to explore their potential in financial tagging tasks. \cite{khatuya2024parameter} proposed the FLAN-FinXC framework, which reformulated the extreme multi-label tagging problem as a generative task. \cite{han2024xbrl} developed the XBRL-Agent system to enhance LLMs’ comprehension of financial reports.
\textbf{(2) Re-Ranking with LLMs:} Recently, LLM-based re-ranking methods have achieved strong results in cross-lingual retrieval, multi-passage ranking, and contextual modeling. For example, \cite{ma2023zero} proposed the unsupervised listwise method LRL, improving cross-lingual retrieval without fine-tuning. \cite{sun2023chatgpt} used fine-tuned LLMs to re-rank paragraph-level texts by relevance. \cite{pradeep2023rankvicuna} introduced RankVicuna for zero-shot listwise document re-ranking. \cite{qin2023large} proposed the Pairwise Ranking Prompting (PRP) method to enhance LLM ranking via pairwise comparison. \cite{chen2024attention} presented the ICR method, which uses attention signals for contextual re-ranking and eases the efficiency bottleneck of generative models.

\section{METHODOLOGY}
In this section, we present the XBRLTagRec framework for predicting XBRL numeral tags, as illustrated in Figure~\ref{fig:overview}. 
Descriptions of these three components are provided below.

\subsection{Tag Document Generation}

We adopt the instruction-tuned FLAN-T5-Large model~\cite{chung2024scaling} as our core generator for text generation under an instruction learning paradigm. In this process, we use parameter-efficient LoRA~\cite{hu2021lora} fine-tuning to FLAN-T5-Large using the training set, significantly reducing training costs while enhancing the model’s adaptability to specific financial tasks. Each training sample consists of three components: an instruction prompt (denoted as \( P \)), a financial statement text \( D_i \), and a question \( Q_x^i \) targeting a specific numeral in the statement. These three components are concatenated to form the final input:
\begin{equation}
\text{D}_x^i = P \oplus D_i \oplus Q_x^i
\end{equation}
where $i$ denotes the $i$-th financial report, $x$ denotes a specific numeral within the financial statement text $D_i$, and $\oplus$ denotes a newline-based joining operation.

The objective of this process is to train the model to generate the corresponding XBRL tag document for the target numeral, marked as $\text{genTagDoc}_x^i$, where $\text{LLM}_{\text{LoRA}}$ denotes the FLAN-T5-Large model fine-tuned with the LoRA technique.
\begin{equation}
\text{genTagDoc}_x^i = \text{LLM}_{\text{LoRA}}(D_x^i)
\end{equation}
During training, we adopt an autoregressive approach to generate the tag document step by step, using the cross-entropy loss as the optimization objective. At each time step, the model generates the next token based on the current input state until the tag document is produced. After training, the fine-tuned model is used during the testing phase, where the input format remains the same as in the training phase, and the output is the generated tag document $\text{genTagDoc}_x^i$. 

It is important that we generate XBRL tag documents rather than individual XBRL tags. Although there exists a one-to-one correspondence between the two, tag documents provide richer semantic structures, enabling finer-grained semantic distinctions and enhancing the model's discriminative ability in extreme classification tasks. Meanwhile, the introduction of the LoRA fine-tuning mechanism avoids full-parameter updates and greatly reduces resource use during training.

\subsection{Semantic Similarity Retrieval}

We adopt the pretrained model~\cite{han2021pre} Sentence-T5-XXL~\cite{ni2021sentence} as the embedding encoder to encode each generated tag document $\text{genTagDoc}_x^i$ produced by the LoRA-tuned FLAN-T5-Large LLM. Similarly, all ground-truth XBRL tag documents are also encoded into their corresponding embedding vectors. After obtaining all embedding vectors, we compute the cosine similarity between $\text{genTagDoc}_x^i$ and each ground-truth XBRL tag document embedding. We then return the Top-10 XBRL tag document set with the highest similarity scores.

\subsection{LLM Zero-Shot Re-Ranking}

To further improve the accuracy of tag prediction, we introduce the LLM Zero-Shot Re-Ranking mechanism. We employ ChatGPT-3.5 (API version is gpt-3.5-turbo-1106) as a re-ranker to refine the Top-10 candidate XBRL tag documents obtained in the previous stage with Sentence-T5-XXL. We select the XBRL tag document that is semantically closest to $\text{genTagDoc}_x^i$ as the final predicted document $\text{predTagDoc}_x^i$, and thus derive the predicted XBRL tag. To fully exploit the semantic information in candidate tag documents and mitigate local biases in the ranking process, we design a multi-round iterative re-ranking strategy that combines \textit{filtering} and \textit{most-frequent selection}. The detailed procedure is as follows:

\textbf{(a) Filtering Strategy:} First, we randomly shuffle the Top-10 candidate documents while preserving their original similarity-based ranking order within each group. The shuffled candidates are equally split into two groups, each containing 5 documents. Next, the target XBRL tag documents of each group and the generated tag document $\text{genTagDoc}_x^i$ are jointly fed into the ChatGPT-3.5 re-ranking module~\cite{sun2023chatgpt}. This module re-ranks the target tag documents based on their relevance to $\text{genTagDoc}_x^i$. 
The target XBRL tag document with the highest relevance in the ranking results is then returned as the most relevant candidate in the group.

\textbf{(b) Multi-Round Iteration:} We repeat the filtering process for several rounds. In each round, one optimal document is selected from each of the two groups, and the selected documents are added to a candidate pool for final selection. Since each iteration adopts a different combination of candidates, this mechanism effectively enhances semantic diversity and mitigates local biases in the ranking process, thereby improving the re-ranking module's robustness and global coverage.

\textbf{(c) Most-Frequent Selection:} We count the cumulative frequency with which each document is selected as the top-ranked candidate across all iterations, denoted as \( f(d) \), whose computation is defined in Eq.\ref{freq}:
\vspace{-6pt}
\begin{equation} \label{freq}
\text{f(d)} = \sum_{t=1}^{T} \mathbb{I}[d = d_t^*]
\end{equation}
where \( T \) denotes the number of iterations, \( d_t^* \) denotes the top-ranked document in iteration \( t \), and \( \mathbb{I}[\cdot] \) is an indicator function.

\begin{figure}[htbp!]
        \centering
        \includegraphics[width=0.49\textwidth]{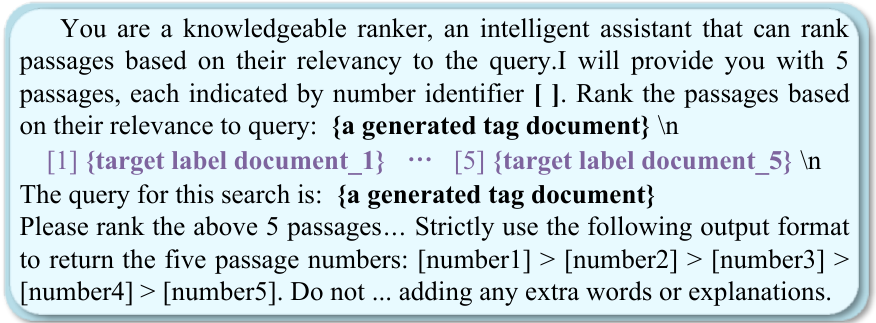}
        \vspace{-20pt} 
        \caption{The prompt guides the LLM to rank the five target label documents by their relevance to the generated tag document and the target label documents.}
        \label{fig:prompt}
\vspace{-0.3em}
\end{figure}

Finally, we select the document with the highest frequency \( f(d) \) as the final tag document prediction $\text{predTagDoc}_x^i$ for the generated tag document $\text{genTagDoc}_x^i$. 
In addition, during implementation, we design a fixed input format with strong instruction constraints for ChatGPT-3.5 to ensure that it returns rankings in a consistent and structured manner. This prompt is adapted from~\cite{sun2023chatgpt}, which is selected due to its strong performance in prior experiments. Based on this, we further refine the original prompt to develop the high-quality re-ranking prompt used in this study. The details are shown in Figure~\ref{fig:prompt}. Overall, the complete algorithmic workflow of the XBRLTagRec framework is shown in Algorithm~\ref{alg:algorithm1}. 

\vspace{-1.2em}

\section{EXPERIMENTAL SETUP} 

\subsection{Dataset}

The FNXL dataset was constructed by~\cite{sharma2023financial} based on publicly available SEC annual 10-K filings. It covers 2,339 publicly listed companies and contains 79,088 sentences along with 142,922 manually annotated financial numeral instances. Each numeral is associated with a corresponding GAAP metric and assigned an appropriate XBRL tag, resulting in a total of 2,794 distinct tags. Notably, the tag distribution exhibits a pronounced long-tail characteristic.
\textit{Since the full FNXL dataset~\cite{sharma2023financial} is not publicly available, we conduct experiments on the partial FNXL dataset released by Khatuya et al.~\cite{khatuya2024parameter}. This subset is derived from the same original dataset and represents the publicly accessible portion used in our experiments.}

\begin{algorithm}[htbp!]
  \footnotesize
  \caption{The process for XBRLTagRec framework} 
  \label{alg:algorithm1}
  \begin{algorithmic}[0.1]
    \REQUIRE
       (1) Financial report: $D_i$, $i\in[1,M]$; (2) Instruction prompt: $P$; (3) Target Numeral: $x$; (4) Question: $Q_x^i$, $i\in[1,M]$ (5) Ground-truth XBRL tag documents: $\text{trueTagDoc}_x^i$, $i\in[1,M]$
    \ENSURE
        The predicted XBRL tag document: $\text{predTagDoc}_x^i$;
        \FOR{$i = 1$; $i \leq M$; $i++$ }
            \STATE $D_x^i = P \oplus D_i \oplus Q_x^i$, where $\oplus$ denotes newline concatenation;
        \ENDFOR
        \FOR{$i = 1$; $i \leq M$; $i++$ }%
            \STATE Generates tag documents: $\text{genTagDoc}_x^i = \text{LLM}_{\text{LoRA}}(D_x^i)$;
        \ENDFOR
        \FOR{$i = 1$; $i \leq M$; $i++$}
            \STATE Compute embeddings vectors: $\mathbf{Vgen}_x^i = \text{Enc}(\text{genTagDoc}_x^i)$, $\mathbf{Vtrue}_x^i = \text{Enc}(\text{trueTagDoc}_x^i)$;
        \ENDFOR
    \STATE $\text{Top10\_Docs} = \text{TopK}_{k=10}\left(\cos(Vgen_x^i, Vtrue_x^i)\right)$
    \STATE Initialize vote counter $f: \text{Top10\_Docs} \rightarrow \mathbb{N}_0$
    \FOR{$t = 1$; $t \leq T$; $t++$ }
        \STATE Randomly divide Top-10 tag documents into two groups $G_1$ and $G_2$, each with 5 documents, \textit{while preserving intra-group ranking order}
        \FOR{$k = 1$; $k \leq 2$; $k++$ }
            \STATE Input $G_k$ and $\text{genTagDoc}_x^i$ into ChatGPT-3.5 to obtain ranking
            \STATE Select top-ranked tag document $d_t^k$
        \ENDFOR
        \STATE Update frequency count: $f(d_t^k) \leftarrow f(d_t^k) + 1$, $k\in[1,2]$
    \ENDFOR
    \STATE Select the document with the highest vote count: $\text{predTagDoc}_x^i = \arg\max_{d} f(d)$
  \end{algorithmic}
\end{algorithm}
\vspace{-1em}

\subsection{Evaluation Metrics}

We employ four evaluation metrics for XBRL tag prediction. \textbf{Macro-Precision} (M-P): computes precision for each class and averages the results. \textbf{Macro-Recall} (M-R): averages recall across all classes. \textbf{Macro-F1} (M-F1): balances precision and recall and is suitable for class-imbalance scenarios~\cite{hinojosa2024performance}. 
In addition, to reflect practical utility and support expert tag recommendation, we introduce \textbf{Hits@1}~\cite{li2025user}.


\subsection{Baseline Methods} 
To confirm the effectiveness of our framework, we compare its performance against the following baselines:
\textbf{(1) FLAN-FinXC}~\cite{khatuya2024parameter}: A strong baseline method that achieves notable performance on extreme financial number labeling tasks. It is built on a generative model and optimized through instruction tuning. \textbf{(2) AttentionXML Pipeline}~\cite{sharma2023financial}: This method builds a BERT-based sequence-to-sequence tagger and models the tagging task as an extreme multi-label classification problem. \textbf{(3) FiNER}~\cite{loukas2022finer}: This method formulates the tagging task as an NER problem, but only covers the 139 most frequent XBRL tags. \textbf{(4) Label Semantics}~\cite{ma2022label}: This method incorporates semantic descriptions of labels and reformulates the tagging task as a standard NER task with entity descriptions as input. \textbf{(5) GalaXC}~\cite{saini2021galaxc}: This method performs joint learning over a constructed document–label graph, effectively integrating label meta-information into the model’s representation.

\subsection{Implementation}
We implement our framework on a 32GB Tesla V100 GPU and use the pretrained checkpoints of FLAN-T5-Large and Sentence-T5-XXL from Hugging Face. During the LoRA fine-tuning process based on the FLAN-T5-Large model, we set the rank to 2, the scaling factor (lora\_alpha) to 32, the dropout rate to 0.05, the learning rate to 5e-4, the number of training epochs to 5, the maximum length of input sequence to 128, and batch size to 8. In the re-ranking module using ChatGPT-Turbo-1106, we divide the Top-10 candidate XBRL tag documents into two groups, each containing 5 documents (group size = 5). In each iteration, the top-ranked tag document from each group is selected and added to a candidate pool for final filtering. This iterative procedure is repeated for 13 rounds.


\subsection{Research Questions}


\textbf{RQ1:} Does our proposed XBRLTagRec framework outperform baseline methods?

\textbf{RQ2:} To what extent do the key parameter designs of our framework affect its overall performance?

\textbf{RQ3:} Does our framework demonstrate generalization and robust performance across various LLMs?


\section{RESULTS}

\subsection{\textbf{RQ1: Does our proposed XBRLTagRec framework outperform baseline methods?}}
To address RQ1, we conduct comparative evaluations of several existing methods and the proposed framework XBRLTagRec on the FNXL dataset. The results are presented in Table~\ref{tab:baseline}. XBRLTagRec achieves the best performance across all evaluation metrics, namely Hits@1, M-P, M-R, and M-F1, highlighting its advantage in the XBRL tagging task.

\vspace{-1em}
\begin{table}[htbp]
    \centering
    \footnotesize
    \setlength\tabcolsep{8pt}
    \renewcommand{\arraystretch}{0.9}
    \caption{Performance evaluation based on Macro and Hits@1 metrics. The best result is shown in bold.}
    \label{tab:baseline}
    \begin{tabular}{cccccccc}
        \toprule
        \multirow{2}{*}{\textbf{Method} } & \multicolumn{4}{c}{\textbf{Eval Metrics}}\\
         \cmidrule{2-5}
         & Hits@1 & M-P & M-R & M-F1 \\
         \midrule
         GalaXC & 0.3596 & 0.2088 & 0.2193 & 0.1870 \\
         AttentionXML &0.7676& 0.5069 &0.4851&0.4754 \\
         Label Semantics & 0.5815 & 0.3360 & 0.3135 & 0.3058 \\
         FiNER & 0.6736 & 0.4104 & 0.4140 & 0.3896 \\
         FLAN-FinXC & 0.8354 & 0.4724 & 0.4881 & 0.4616 \\
         $\textbf{XBRLTagRec}_{(\textbf{1-2 iters})}$ & 0.8526 & 0.4936 & 0.5018 & 0.4777 \\
         $\textbf{XBRLTagRec}_{(\textbf{1-8 iters})}$ & \textbf{0.8618} & \textbf{0.5171} & \textbf{0.5323} & \textbf{0.5050} \\
         \bottomrule
    \end{tabular}
\vspace{-1em}
\end{table}

Specifically, compared with existing baselines, XBRLTagRec already surpasses all baseline methods when the iteration range is limited to 1–2 rounds. When the range is extended to 1–8 rounds, the performance of XBRLTagRec further improves, with Hits@1 increasing to 0.8618. Relative to the strongest baseline, FLAN-FinXC, Hits@1, M-P, M-R, and M-F1 improve by 2.64\%, 4.47\%, 4.42\%, and 4.34\%, respectively. These improvements demonstrate that XBRLTagRec not only achieves higher accuracy in top-ranked label prediction but also attains a more balanced enhancement between precision and recall, leading to a substantial improvement in overall F1 performance. In comparison with other baselines such as FiNER, the improvements are even more pronounced; for example, relative to FiNER, Hits@1 increases by 18.82\% and M-F1 by 11.54\%, further validating the effectiveness of the proposed framework for XBRL tagging. In summary, XBRLTagRec outperforms all baselines on the FNXL dataset, underscoring its strong practical value and potential for adoption in financial XBRL tagging tasks.

\vspace{-0.6em}
\begin{table}[htbp]
\centering
\small
\begin{tabular}{|p{8.5cm}|}
\hline 
\rowcolor{blue!5} \textbf{Answer for RQ1}: \textit{Our XBRLTagRec framework outperforms all baselines in the XBRL tagging task within the financial domain.} \\
\hline 
\end{tabular} 
\vspace{-1.5em}
\end{table}

\subsection{\textbf{RQ2: To what extent do the key parameter designs of our framework affect its overall performance?}}
To answer RQ2, we conduct three empirical studies focusing on three core parameters: the iteration range, the group size, and the group ordering strategy. The results are presented in Table~\ref{tab:model_iters}-Table~\ref{tab:order}, respectively. We then perform a comparative analysis based on the evaluation metrics.

\textbf{(1) Impact of different iteration ranges on model performance:}
Table~\ref{tab:model_iters} shows the tag prediction performance of XBRLTagRec under different iteration settings. The performance shows an ``initial increase followed by stability" trend as iterations increase. Hits@1 reaches its highest value at iterations 1-13, while M-P, M-R, and M-F1 achieve their best values at iterations 1-8. The average values of the four metrics (denoted as avg) indicate that iterations 1-8 provide the optimal overall performance, balancing performance improvement with computational cost. After the 8th iteration, the performance improvement slows, and metrics such as M-F1 decline. This suggests that excessive iterations may introduce redundant information, reducing the discriminative power of the re-ranking process. Therefore, selecting an appropriate number of iterations can enhance performance, prevent overfitting, and avoid unnecessary computation. The results confirm that the iteration selection mechanism of XBRLTagRec is stable, convergent, and scalable under reasonable settings.

\vspace{-1em}
\begin{table}[htbp]
    \centering
    \footnotesize
    \setlength\tabcolsep{6pt}
    \renewcommand{\arraystretch}{0.9}
    \caption{The impact of iteration range on XBRLTagRec's re-ranking module. The best result is shown in bold.}
    \label{tab:model_iters}
    \begin{tabular}{cccccccc}
        \toprule
        \multirow{2}{*}{\textbf{Iteration Range} } & \multicolumn{4}{c}{\textbf{Eval Metrics}}\\
         \cmidrule{2-6}
         & Hits@1 & M-P & M-R & M-F1 & AVG \\
         \midrule
         1-2 iters & 0.8526 & 0.4936 & 0.5018 & 0.4777 & 0.5814 \\
         1-3 iters & 0.8543 & 0.5028 & 0.5140 & 0.4887 & 0.5900 \\ 
         1-4 iters & 0.8581 & 0.5089 & 0.5193 & 0.4947 & 0.5953 \\ 
         1-5 iters & 0.8599 & 0.5112 & 0.5274 & 0.5002 & 0.5997 \\ 
         1-6 iters & 0.8619 & 0.5157 & 0.5305 & 0.5039 & 0.6030 \\ 
         1-7 iters & 0.8624 & 0.5146 & 0.5304 & 0.5032 & 0.6027 \\ 
         1-8 iters & 0.8618 & \textbf{0.5171} & \textbf{0.5323} & \textbf{0.5050} & \textbf{0.6041} \\ 
         1-9 iters & 0.8634 & 0.5110 & 0.5300 & 0.5018 & 0.6016 \\ 
         1-10 iters & 0.8634 & 0.5126 & 0.5312 & 0.5026 & 0.6025 \\ 
         1-11 iters & 0.8634 & 0.5114 & 0.5315 & 0.5025 & 0.6022 \\ 
         1-12 iters & 0.8636 & 0.5097 & 0.5299 & 0.5010 & 0.6011 \\ 
         1-13 iters & \textbf{0.8640} & 0.5099 & 0.5287 & 0.5005 & 0.6008\\ 
         \bottomrule
    \end{tabular}
    \vspace{-1em}
\end{table}

\textbf{(2) Impact of different group sizes on model performance:}
Table~\ref{tab:group} reports the results of XBRLTagRec on the FNXL dataset under different group size settings. The experimental results show that group size has a significant impact on performance: when it increases from 3 to 5, all evaluation metrics improve, indicating that a larger candidate group can enhance the model’s discriminative capability and decision space. However, when the group size is further expanded to 6 or 7, although Hits@1 increases slightly, metrics such as M-F1 exhibit degradation and the average score no longer improves. This suggests that an excessively large group size yields only marginal gains and may even introduce distracting documents. Therefore, setting the group size to 5 achieves the best balance between performance and computational cost, further confirming that an appropriate group size in the re-ranking stage is a key factor influencing overall effectiveness.

\vspace{-1.5em}
\begin{table}[htbp]
    \centering
    \footnotesize
    \setlength\tabcolsep{7pt}
    \renewcommand{\arraystretch}{0.9}
    \caption{The impact of group size on the XBRLTagRec re-ranking module. The best result is shown in bold.}
    \label{tab:group}
    \begin{tabular}{cccccccc}
        \toprule
        \multirow{2}{*}{\textbf{Group Size} } & \multicolumn{5}{c}{\textbf{Eval Metrics}}\\
         \cmidrule{2-6}
         & Hits@1 & M-P & M-R & M-F1 & AVG \\
         \midrule
         3 & 0.8500 & 0.5023 & 0.5167 & 0.4903 & 0.5898 \\
         4 & 0.8601 & 0.5056 & 0.5209 & 0.4944 & 0.5953 \\
         5 & 0.8618 & \textbf{0.5171} & \textbf{0.5323} & \textbf{0.5050} & \textbf{0.6041} \\
         6 & \textbf{0.8783} & 0.5116 & 0.5262 & 0.4997 & 0.6040 \\
         7 & 0.8733 & 0.5117 & 0.5263 & 0.4985 & 0.6025 \\
         \bottomrule
    \end{tabular}
\vspace{-1em}
\end{table}

\textbf{(3) Impact of different grouping strategies on model performance:}
Table~\ref{tab:order} presents the performance of XBRLTagRec under two grouping strategies: Order-Preserving, which maintains the original cosine similarity ranking, and Order-Shuffled, which applies random sorting. The results indicate that Order-Preserving significantly outperforms Order-Shuffled, achieving higher scores across all metrics. For instance, Order-Shuffled drops to 0.4663 in M-F1, and avg decreases by approximately 3\%. This suggests that keeping the cosine similarity ranking helps the model make more accurate re-ranking decisions, while random sorting reduces the effectiveness of re-ranking. These experimental results demonstrate that properly preserving the initial semantic ranking contributes to improving ranking accuracy and robustness.

\vspace{-1.5em}
\begin{table}[htbp]
    \centering
    \footnotesize
    \setlength\tabcolsep{5.5pt}
    \renewcommand{\arraystretch}{0.9}
    \caption{The impact of grouping strategies in the XBRLTagRec's re-ranking module. The best result is highlighted in bold.}
    \label{tab:order}
    \begin{tabular}{cccccccc}
        \toprule
        \multirow{2}{*}{\textbf{Grouping Strategy} } & \multicolumn{4}{c}{\textbf{Eval Metrics}}\\
         \cmidrule{2-6}
         & Hits@1 & M-P & M-R & M-F1 & AVG \\
         \midrule
         Order-Preserving & \textbf{0.8618} & \textbf{0.5171} & \textbf{0.5323} & \textbf{0.5050} & \textbf{0.6041} \\
         Order-Shuffled & 0.8557 & 0.4900 & 0.4871 & 0.4663 & 0.5748 \\
         \bottomrule
    \end{tabular}
    \vspace{-1em}
\end{table}

In summary, to use the re-ranking module, we adopt this configuration for XBRLTagRec: 1–8 iterations to balance performance and cost, a group size of 5 to control ranking interference, and an Order-Preserving strategy to keep cosine-based ranking consistency. This setting yields stable performance across experiments and offers guidance for future tasks.

\vspace{-0.6em}
\begin{table}[htbp]
\centering
\small
\begin{tabular}{|p{8.5cm}|}
\hline 
\rowcolor{blue!5} \textbf{Answer for RQ2}: \textit{Three key parameters—iteration range, group size, and grouping strategy—are crucial for performance gains.} \\
\hline 
\end{tabular} 
\end{table}
\vspace{-1.6em}

\subsection{\textbf{RQ3: Does our framework demonstrate generalization and robust performance across various LLMs?}}
To answer RQ3, we replace the LLM in the re-ranking module under a unified generation and retrieval pipeline, selecting four mainstream large-scale models—ChatGPT-3.5-Turbo-1106~\cite{achiam2023gpt}, GPT-4o~\cite{achiam2023gpt}, DeepSeek-V3~\cite{liu2024deepseek}, and ERNIE-3.5-8K~\cite{baidu2025ernie35}—as re-rankers to evaluate their performance on the FNXL dataset. The detailed results are presented in Table~\ref{tab:differ}. 


\vspace{-0.1em}
\begin{table}[htbp]
    \centering
    \footnotesize
    \setlength{\textfloatsep}{-19pt}
    \setlength\tabcolsep{4pt}
    \renewcommand{\arraystretch}{0.9}
    \caption{Performance of XBRLTagRec with different LLM-based re-rankers (Evaluated with 1–2 iterations)}
    \begin{tabular}{cccccccc}
        \toprule
        \multirow{2}{*}{\textbf{Model+LLM} } & \multicolumn{4}{c}{\textbf{Eval Metrics}}\\
         \cmidrule{2-6}
         & Hits@1 & M-P & M-R & M-F1 & AVG \\
         \midrule
         FLAN-FinXC (SOTA) & 0.8354 & 0.4724 & 0.4881 & 0.4616 & 0.5644\\
         \midrule
         $\text{XBRLTagRec}_{+\text{ChatGPT-3.5}}$ & 0.8526 & 0.4936 & 0.5018 & 0.4777 & 0.5814\\
         $\text{XBRLTagRec}_{+\text{DeepSeek-V3}}$ & 0.8672 & 0.4856 & 0.4935 & 0.4706 & 0.5792\\
         $\text{XBRLTagRec}_{+\text{ERNIE-3.5-8K}}$  & 0.8594 & 0.4992 & 0.5028 & 0.4827 & 0.5860\\
         $\text{XBRLTagRec}_{+\text{GPT-4}}$  & 0.8730 & 0.5160 & 0.5227 & 0.4998 & 0.6029\\
         \bottomrule
         \label{tab:differ}
    \end{tabular}
    \vspace{-3em}
\end{table}

As shown in Table~\ref{tab:differ}, the XBRLTagRec framework maintains stable and strong performance across all four models, demonstrating good model-independence and cross-model adaptability. Although discrepancies still exist among the models, the overall variance remains within a controllable range, indicating that the framework exhibits strong robustness under model-replacement scenarios.
Specifically, GPT-4 achieves the best performance across all metrics; ERNIE-3.5-8K ranks second to GPT-4 in terms of M-F1, making it more suitable for scenarios that require a balance between precision and recall; DeepSeek-V3 attains a higher Hits@1 but the lowest M-F1, reflecting its bias toward top-ranking optimization, although ChatGPT-3.5 has the lowest Hits@1, its M-F1 exceeds that of DeepSeek-V3, indicating a more stable ranking capability. Considering both overall performance and computational cost, this study ultimately selects ChatGPT-3.5 as the primary experimental model.
In summary, the experiment keeps the core structure of the framework unchanged and only replaces the LLM in the re-ranking module. The results show that the framework produces stable and reliable outcomes across different models, highlighting its modular and plug-and-play design advantages, as well as demonstrating good scalability that supports convenient integration of more types of LLMs.

\vspace{-1em}
\begin{table}[htbp]
\centering
\small
\begin{tabular}{|p{8.38cm}|}
\hline 
\rowcolor{blue!5} \textbf{Answer for RQ3}: \textit{Considering computational cost, XBRLTagRec outperforms baselines with only two iterations across LLMs.} \\
\hline 
\end{tabular} 
\end{table}
\vspace{-0.9em}

\section{CONCLUSION}
In conclusion, we propose XBRLTagRec, an end-to-end framework for automated XBRL tag matching in financial texts. By integrating instruction-tuned LLMs, semantic retrieval, and iterative re-ranking, XBRLTagRec effectively addresses semantic ambiguity and contextual complexity in large-scale tag systems. Experimental results demonstrate clear improvements over existing methods, including the state-of-the-art FLAN-FinXC. Its modular design further ensures flexibility and scalability, making XBRLTagRec a practical and robust solution for financial data annotation.

\section{LIMITATIONS}
This study has three limitations: (1) iterative re-ranking with ChatGPT-3.5 increases computational and API costs; (2) the impact of different label embedding models is not examined; (3) zero-shot numeral labeling using untuned general LLMs is not investigated. Future work will explore efficient zero-shot methods and external knowledge integration.

\section*{Acknowledgments}
The authors thank the General Program of Applied Basic Research of Yunnan Province (No.202301AT070184), the Open Project Program of Yunnan Key Laboratory of Intelligent Systems and Computing (No.ISC22Y08), and the Open Research Project of the Yunnan University (YNU) Resilience and Excellence Children's Character Development Platform (NO.K207003250006).

\bibliographystyle{IEEEtran}
\bibliography{reference}
\end{document}